\RequirePackage{graphicx}
\documentclass[aps,prl,twocolumn,showpacs,10pt,longbibliography]{revtex4-2}
\usepackage{amssymb}
\usepackage{enumitem}
\usepackage{mathtools}
\usepackage{xcolor}
\usepackage{float}
\usepackage{tikz}
\usetikzlibrary{quantikz}
\usepackage{natbib}
\usepackage{graphicx}
\usepackage{dcolumn}
\usepackage{bm}
\usepackage{mathtools}
\usepackage{braket}
\usepackage{algorithm}
\usepackage{algpseudocode}
\usepackage[utf8x]{inputenc}
\usepackage{amsmath}
\usepackage{multirow}
\usepackage{amsfonts}
\usepackage{amssymb}
\usepackage{rotating} 

\usepackage{amsthm}
\theoremstyle{plain}
\theoremstyle{definition}

\newcommand{\Gam}[4]{\hat{a}^\dagger_{#1} \hat{a}^\dagger_{#2} \hat{a}^{}_{#3} \hat{a}^{}_{#4}}

\begin{document}

\title{Verifiably Exact Solution of the Electronic Schr{\"o}dinger Equation on Quantum Devices}





\author{Scott E. Smart and David A. Mazziotti}
\email[]{damazz@uchicago.edu}
\affiliation{Department of Chemistry and The James Franck Institute, The University of Chicago, Chicago, IL 60637}%
\date{Submitted February 25, 2023}

\begin{abstract}
Quantum computers have the potential for an exponential speedup of classical molecular computations. However, existing algorithms have limitations; quantum phase estimation (QPE) algorithms are intractable on current hardware while variational quantum eigensolvers (VQE) are dependent upon approximate wave functions without guaranteed convergence.  In this Article we present an algorithm that yields verifiably exact solutions of the many-electron Schr{\"o}dinger equation.  Rather than solve the Schr{\"o}dinger equation directly, we solve its contraction over all electrons except two, known as the contracted Schr{\"o}dinger equation (CSE).  The CSE generates an exact wave function ansatz, constructed from a product of two-body-based non-unitary transformations, that scales polynomially with molecular size and hence, provides a potentially exponential acceleration of classical molecular electronic structure calculations on ideal quantum devices.  We demonstrate the algorithm on both quantum simulators and noisy quantum computers with applications to H$_{2}$ dissociation and the rectangle-to-square transition in H$_{4}$.  The CSE quantum algorithm, which is a type of contracted quantum eigensolver (CQE), provides a significant step towards realizing verifiably accurate but scalable molecular simulations on quantum devices.
\end{abstract}

\maketitle


\maketitle

\section{Main}

The wave function of a many-particle quantum system scales exponentially with the number of particles, and hence, it can be computationally advantageous to compute the energy and other one-and two-particle expectation values (i.e. the two-particle reduced density matrix (2-RDM)) without computing the wave function~\cite{Mazziotti2007_book, Coleman2000, Cioslowski2000, Mazziotti2012rev, Coleman1963, Harriman1984}. A natural equation for the determining the 2-RDM is obtained from contracting the matrix formulation of the Schr{\"o}dinger equation over all particles save two, generating the contracted Schr{\"o}dinger equation (CSE)~\cite{Mazziotti1998, Nakatsuji1996, Colmenero1993}.  The CSE has the significant, non-trivial property that a wave function satisfies the CSE if and only if it satisfies the Schr{\"o}dinger equation\cite{Mazziotti1998, Nakatsuji1976}.  In classical electronic structure algorithms, however, the CSE is indeterminate in the solution of the 2-RDM because it depends upon not only the 2-RDM but also the higher three- and four-particle RDMs (3- and 4-RDMs).  While the indeterminacy of the equation can be removed by reconstructing higher RDMs in terms of the 2-RDM,\cite{Mazziotti1998_cumulants, Mazziotti1999, Misiewicz2020, Herbert2002}, the reconstruction introduces an approximation, which limits applications of the CSE from being exact.

In this Article we show that many-particle quantum systems can be solved at non-exponential cost on a noiseless quantum computer from the solution of the CSE for the 2-RDM.  The reconstruction of higher RDMs in the CSE, required on classical computers, can be avoided on a quantum computer through a combination of state preparations and 2-RDM measurements.  The CSE extends a family of algorithms, known as contracted quantum eigensolvers (CQE), in which the residual of a contracted eigenvalue equation is solved on a quantum computer to generate a solution of the original eigenvalue problem~\cite{Smart2021_PRL, Smart2022_PRA, Smart2022_benzyne, Smart2022_JCTC, Mazziotti2021}.  The solution of the CSE generates a series of non-unitary, two-body transformations that provide an exact ansatz for the wave function~\cite{Mazziotti2004, Mazziotti2020, Nakatsuji2001}. The wave function ansatz is scalable because at each iteration the transformation has only two-body terms to check satisfaction of the CSE regardless of the number of electrons and orbitals.  In all previous versions of the CQE, the anti-Hermitian part of CSE (ACSE), which keeps the two-body exponential transformations unitary, was used as the contracted equation.\cite{Mazziotti2007,Mazziotti2006}.  However, the ACSE, unlike the CSE, does not necessarily satisfy the Schr{\"o}dinger equation upon its solution, which can in principle lead to problems where the ACSE is satisfied away from a eigenstate.

 In contrast to many variational quantum eigensolvers (VQE)~\cite{Peruzzo2014, Kandala2017, Tilly2021} which use approximate wave function ans{\"a}tze, the proposed CQE for solving the CSE converges in the absence of noise to a solution of the CSE that corresponds to an exact many-particle solution of the Schr{\"o}dinger equation. The CSE ansatz for the wave function is more scalable than the conventional coupled cluster ansatz because it is exact with only two-body transformations~\cite{Mazziotti2004} while the coupled cluster ansatz requires 2-to-$N$-particle excitations~\cite{Romero2017, Mazziotti2004}. Additionally, the CQE approach performs the optimization based on the residual of the contracted equation, and hence, does not require any VQE-based subroutines, seen in adaptive or iterative based approaches~\cite{Lee2018, Nakatsuji2001, Grimsley2018, Liu2022}. In this work we provide examples of the exactness of the CSE-based CQE, and additionally apply the CQE to molecular H$_{2}$ on a superconducting quantum device, as well as noiseless simulations of the rectangle-to-square transition in H$_{4}$ with quantum simulators.


For many-particle quantum systems with at most pairwise interactions, consider the contraction of the matrix formulation of the Schr{\"o}dinger equation to generate the CSE\cite{Mazziotti1998, Mazziotti2004, Nakatsuji1996, Colmenero1993, Nakatsuji1976},
\begin{equation}
 \langle \Psi | \Gam{i}{j}{l}{k} (\hat{H}-E) | \Psi \rangle = 0
\end{equation}
where ${\hat H}$ is the Hamiltonian operator, $| \Psi \rangle$ is the wave function of a given state, $E$ is the energy of the state, and $a^{\dagger}_{i}$($a^{}_{i}$) is the second-quantized creation (annihilation) operator for a particle in orbital $i$.  For convenience we assume that the wave function and the Hamiltonian are real.  The CSE has the following important property: if the Hamiltonian has at most pairwise interactions, a wave function satisfies the CSE if and only if it satisfies the Schr{\"o}dinger equation.  The Schr{\"o}dinger equation clearly implies the CSE, but the opposite direction is provable from showing that the CSE implies the variance which in turn implies the Schr{\"o}dinger equation~\cite{Nakatsuji1976, Mazziotti1998}.  The CSE, it has been shown previously~\cite{Mazziotti2004, Mazziotti2020}, implies that the exact wave function $| \Psi \rangle$ has a minimal parametrization in which it is expressed as a product of two-body exponential transformations applied to a mean-field reference wave function $| \Psi_{0} \rangle$ as follows:
\begin{equation}
| \Psi \rangle = {\rm e}^{{\hat J}_{n}} \cdots {\rm e}^{{\hat J}_{2}} {\rm e}^{{\hat J}_{1}} | \Psi_{0} \rangle
\end{equation}
where ${\hat J}_{n}$ for each $n$ is a general two-body operator
\begin{equation}
    \hat{J}_n = \sum_{pqst}{{}^{2} J^{st;pq}_{n}\Gam{p}{q}{t}{s}} .
\end{equation}
in which ${}^{2} J^{pq;st}_n$ is a two-body matrix element.  Importantly, differentiating the energy with respect to the matrix elements of ${}^{2} J_{n}$ yields the residual of the CSE
\begin{eqnarray}
    {}^{2} R^{ij;kl} & = & \lim_{\epsilon \rightarrow 0} \frac{1}{2\epsilon}\frac{d}{dJ^{ij;kl}_n} \langle \Psi | {\rm e}^{ \epsilon \hat{J}_{n}^{\dagger} } ( \hat{H} - E ) {\rm e}^{\epsilon \hat{J}_{n}} | \Psi  \rangle \\
    & = &  \langle \Psi |  \Gam{i}{j}{l}{k} (\hat{H} - E) | \Psi \rangle .
\end{eqnarray}
Upon convergence of the energy at the $n^{\rm th}$ iteration, the gradient of the energy with respect to the parameters ${}^{2} J^{pq;st}_{n}$ vanishes, implying the satisfaction of the CSE.  Consequently, all local minima of this ansatz for the wave function correspond to solutions of the CSE and hence, solutions of the Schr{\"o}dinger equation, proving that this wave-function ansatz is exact~\cite{Mazziotti2004, Mazziotti2020}.  On classical computers, because the wave function has exponential-scaling cost in terms both computation and storage, the CSE is re-expressed in terms of RDMs\cite{Mazziotti1998, Nakatsuji1996, Colmenero1993, Mazziotti2002} with the higher-order RDMs being approximated as functionals of the 2-RDM~\cite{Mazziotti1998_cumulants}.  On a quantum computer, however, the CSE ansatz for the exact wave function can be prepared and the 2-RDM measured at non-exponential-scaling cost, allowing us to solve the CSE for in principle exact simulations of many-particle quantum systems.

\begin{table}
  \caption{CQE for solving the contracted Schr{\"o}dinger equation by non-unitary two-body exponentials transformations .}
  \label{t:CSE_algorithm}
  \begin{ruledtabular}
  \begin{tabular}{l}
  {\bf CQE Algorithm} \\
  \hspace{0.1in} Set $0 \leftarrow n $ \\
  \hspace{0.1in} {Initialize} $| \Psi_{0} \rangle, \hat{R}_0$ \\
  \hspace{0.1in} {While $||{}^2 R_n || >\delta$}:  \\
  \hspace{0.2in} { {\bf Step 1:} $\min E_n(\hat{J}_n)  $}  \\
  \hspace{0.2in} { {\bf Step 2:} Evaluate ${}^{2} \hat{R}_{n+1},~\hat{J}_{n+1}[{}^2 \hat{R}_{n+1}] $} \\
  \hspace{0.2in} { {\bf Step 3:} Construct $| \psi_{n+1}\rangle = e^{\hat{J}_{n}}|\psi \rangle $ }\\
  \hspace{0.2in} { {\bf Step 3:} }$n+1 \leftarrow n $ \\
  \end{tabular}
  \end{ruledtabular}
\end{table}

We address both of these aspects by separating the CQE into Hermitian and anti-Hermitian parts, denoted as the HCSE and ACSE respectively. First, we can divide the operator $\hat{J}$ into its Hermitian $\hat{J}_H$ and anti-Hermitian $\hat{J}_A$ parts:
\begin{equation}
    {}^2 \hat{J}^H_n = \frac{{}^2\hat{J}_n + {}^2\hat{J}^\dagger_n}{2}, ~~~ {}^2\hat{J}^A_n = \frac{{}^2\hat{J}_n - {}^2\hat{J}^\dagger_n}{2}.
\end{equation}
We can then implement these operators in a trotterized form sequentially:
\begin{equation}
    |\psi_{n+1} \rangle = \frac{1}{\sqrt{N}} e^{\epsilon \hat{J}^{n}_H} e^{\epsilon \hat{J}^n_A} | \psi_n \rangle
\end{equation}
where $N= \langle \psi_{n+1} |\exp{(2 \epsilon\hat{J}^n_H)}|\psi_{n+1} \rangle  $ is the appropriate normalization. To implement the exponential operator $e^{ \epsilon \hat{J}_n^H}$, there are a number of approaches for non-unitary dynamics which can be adapted for the current work.

First, we can project the operator onto the two-body unitary space, similar to the least-squares based imaginary time-evolution technique~\cite{Motta2019}, except that we are projecting both unitary and non-unitary components. This approach generates a linear approximation to the original operator that is unitary but requires measurement of the 4-particle RDM (4-RDM).  The second idea, which we develop here, dilates the wave function using a single ancilla, and applies the first-order non-unitary transformation repeatedly to effect the optimization.

Consider an operator $V[\hat{J}_H^n]$
\begin{equation}
    V[\hat{J}^H_n] = \begin{bmatrix} 1 & \epsilon \hat{J}^H_n\\ - \epsilon \hat{J}^H_n & 1  \end{bmatrix},
\end{equation}
which acts to produce a state:
\begin{equation}
    V[\hat{J}^H_n] \begin{bmatrix} |\psi \rangle & |\psi \rangle \end{bmatrix}^T = \begin{bmatrix} {\rm e}^{\epsilon\hat{J}^H_n}|\psi \rangle  & {\rm e}^{-\epsilon \hat{J}^H_n} | \psi \rangle  \end{bmatrix} + O(\epsilon^2) .
\end{equation}
This can be realized in a quantum algorithm by expanding the state space with a single ancilla, and then applying a Pauli gadget with the Pauli-Y matrix:
\begin{equation}
    V[\hat{J}^H_n] = {\rm e}^{i \delta \hat{Y}_a \otimes \hat{J}_{n}^H}.
\end{equation}
This method works well for small steps, and is similar to techniques for approximating non-unitary time evolution~\cite{Schlimgen2021}. The remaining problem relates to subsequent evolutions, which is not addressed in most work. To proceed exactly, we can either append another ancilla~\cite{Daskin2017} or perform a projective measurement to select the particular ancilla state. Both of these approaches, however, decrease the success probability roughly by $\frac{1}{2}$, having exponentially decreasing success rates~\cite{Kosugi2021}. Instead, we directly implement the next operator without an ancilla or projective measurement, which can be shown to generate the target operator to first order.  After several steps (on the order of $\frac{1}{\epsilon}$) the first-order approximation will fail, and hence, we use a loose Wolfe condition~\cite{Robinson2006} to determine when to perform a reset or dilation onto another ancilla. The proposed method still leads to an exponentially decreasing success probability, but with a much slower decay than in the previous methods. In theory, performing a type of amplitude amplification~\cite{Brassard2002,Nakaji2020} could further mitigate this effect, although for near-term results this is likely not feasible. Details comparing effective strategies for implementing this operator will be compared in future work.

To obtain the residuals, we again can relate the residuals of the CSE, to the residuals of the HCSE and ACSE:
\begin{equation}
    {}^2 R^{ij}_{kl} = \frac{1}{2} ({}^2 S^{ij}_{kl} + {}^2 A^{ij}_{kl}),
\end{equation}
where the residuals of the HCSE and ACSE are respectively defined as:
\begin{eqnarray}\label{eq:hcse}
    {}^2 S^{ij}_{kl} & = & \langle\psi | \{ \Gam{i}{j}{l}{k},\hat{H}-E \}| \psi \rangle , \\
\label{eq:acse}
    {}^2 A^{ij}_{kl} & = & \langle\psi | [ \Gam{i}{j}{l}{k}, \hat{H} ] | \psi \rangle .
\end{eqnarray}
As mentioned above, these require the 4-RDM and 3-RDM respectively, and so obtaining information on the HCSE in particular is challenging. We again turn to a ancilla-based expansion of the wave function, where we are looking to perturb the wave function and recover information on the contracted equation to some controlled approximation.  While there likely are ways to simultaneous recover the ACSE and HCSE residuals, in a circuit-based measurement these are similar to the real and imaginary tomography of a 2-RDM, which naturally can be separated. Hence, we simply separate the ACSE and HCSE residuals into two auxiliary state methods.  This method of obtaining the ACSE to second-order accuracy in $\delta$ has been documented previously,\cite{Smart2021_PRL} and we can envision a method of obtaining the HCSE to a similar accuracy as follows.

\begin{figure}
    \centering
    \includegraphics[scale=0.22]{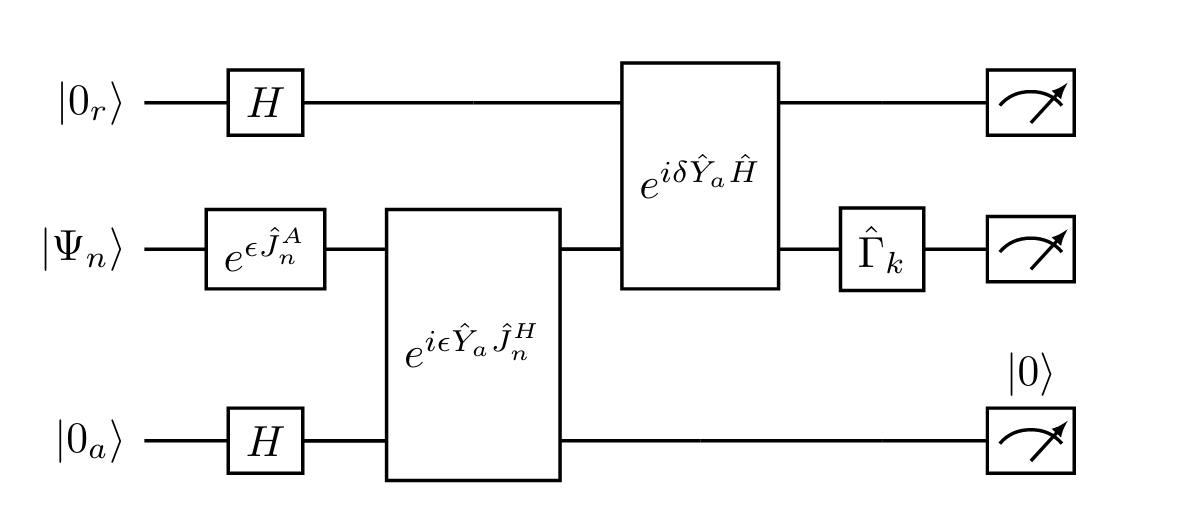}
    \caption{Method of implementing the CSE ansatz for the contracted quantum eigensolver. At a given iteration $n$, we implement the exponential of the anti-Hermitian operator $\hat{J}^A_n$ directly, and then non-unitary exponential of $\hat{J}^H_n$ using a single ancilla qubit. Information on the residuals can be obtained using a conditioned time evolution operator, as well as the traditional time evolution operator. }
    \label{fig:circuit}
\end{figure}

Let $W$ be a unitary acting on a single ancilla coupled with $|\psi\rangle$ to produce an auxiliary state $|\Upsilon \rangle$.
\begin{equation}
    {\hat W} = \begin{bmatrix} 1 - \frac{\delta^2}{2} \hat{H}^2 & +\hat{H} \\ -\hat{H} & 1 - \frac{\delta^2}{2}\hat{H}^2 \end{bmatrix},
\end{equation}
and
\begin{equation}
    | \Upsilon \rangle = \frac{1}{\sqrt{2}} {\hat W} \begin{bmatrix} |\psi\rangle  & |\psi\rangle  \end{bmatrix}^T.
\end{equation}
Performing the following 2-RDM related measurement:
\begin{equation}
 \hat{M} {}^{ij}_{kl} =    \begin{bmatrix} \Gam{i}{j}{l}{k} & 0 \\ 0 & -\Gam{i}{j}{l}{k} \end{bmatrix},
\end{equation}
yields the following residuals:
\begin{equation}
    \frac{1}{\delta}\langle \Upsilon_n | \hat{M} | \Upsilon_n \rangle = {}^2 R^{ij;kl}_n   + O(\delta^2)
\end{equation}
On a quantum computer, we can readily perform this operation where ${\hat W}$ corresponds to a Pauli-conditioned time evolution operator and $\hat{M}^{ij}_{kl}$ to a Pauli-conditioned 2-RDM measurement:
\begin{equation}
    \hat{W} = \exp{(i \delta Y_a \otimes \hat{H})}, ~~\hat{M}^{ij}_{kl} = Z_a \otimes \Gam{i}{j}{l}{k}
\end{equation}
in which $Y_a, Z_a$ are the Pauli Y and Z gates acting on the ancilla $a$ which we have prepared by applying the Hadamard transform to the ancilla. Because the Pauli terms are present in every operator, these Pauli conditioned operators possess the same scaling as their original operators. In practice, care must be taken with the relative precisions such that $\delta$ is significantly larger than the sampling errors. An approach involving high precision measurements using techniques of amplitude estimation would alleviate these concerns~\cite{Brassard2002,Nakaji2020}, or possibly via expansions of derivative-based estimates for operators\cite{Crooks2019}.  Figure~\ref{fig:circuit} shows the implementation of the ansatz for a particular iteration.

\section{Model and Molecular Applications}

We consider three applications: (i) a pairing spin model, (ii) the dissociation of rectangular H$_4$, and (iii) the dissociation of H$_{2}$.  The pairing spin model with two pairs of electrons, whose details are given in the Methods, has three independent degrees of freedom corresponding to the excitation of zero, one, or two pairs that can be mapped to the unit sphere.  Using a variety of initial guesses, we compute the ground-state energy on a noiseless simulator by minimizing the residuals of the ACSE and CSE, respectively. Figure~2 shows the solution trajectories on the unit sphere, generated by the iterations in the algorithms, for the ACSE and CSE for each of the initial guesses.  The ACSE converges for many initial guesses to spurious solutions lying on the equator (shaded in grey) where the ACSE's residual vanishes. In contrast, the two-electron CSE correctly converges for each of the initial guesses to the unique ground state of the four-electron Schr{\"o}dinger equation.  Solution by the ACSE would require a fourth-order excitation ansatz.  These results, which are consistent with Ref.~\cite{Mazziotti2004}, highlight the importance of the CSE rather than the ACSE as the stationary condition for a CQE.

\begin{figure}
    \centering
    \includegraphics[scale=0.2]{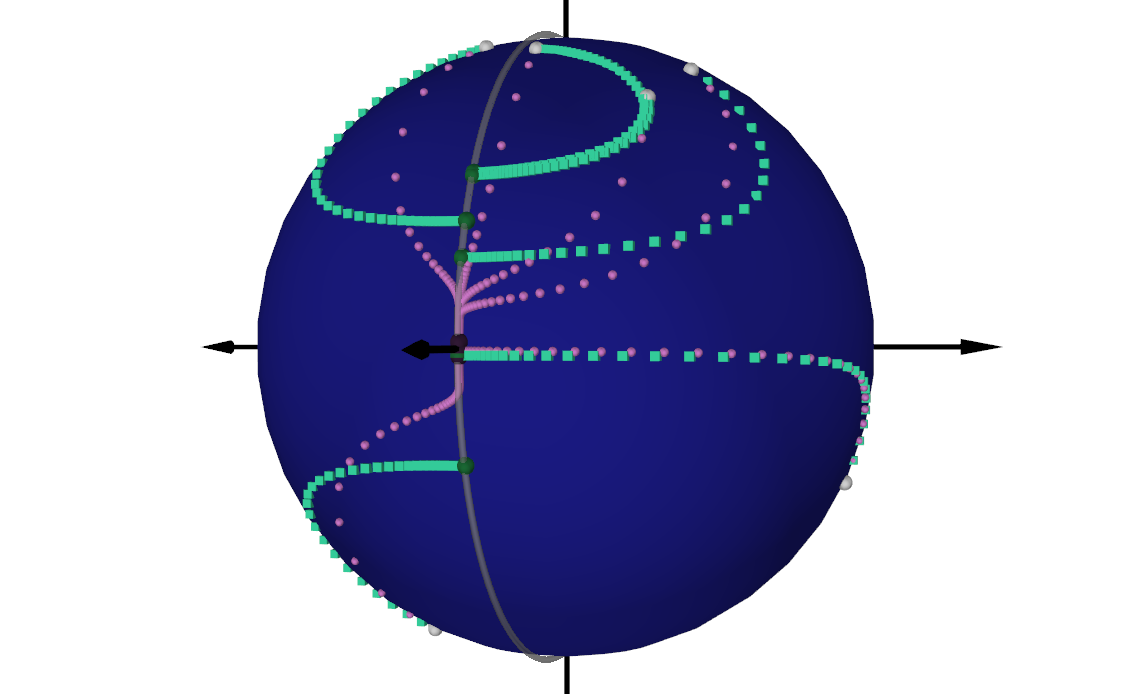}
    \caption{Trajectories for the ACSE and CSE on the unit sphere, generated by the iterations in the algorithms from several initial guesses for a three-state system separated by sequential double excitations (i.e. the ground state and second excited state are separated by quadruple excitations).}
    \label{fig:fig1_acse_cse_comp}
\end{figure}

\begin{figure}
    \centering
    \includegraphics[scale=0.52]{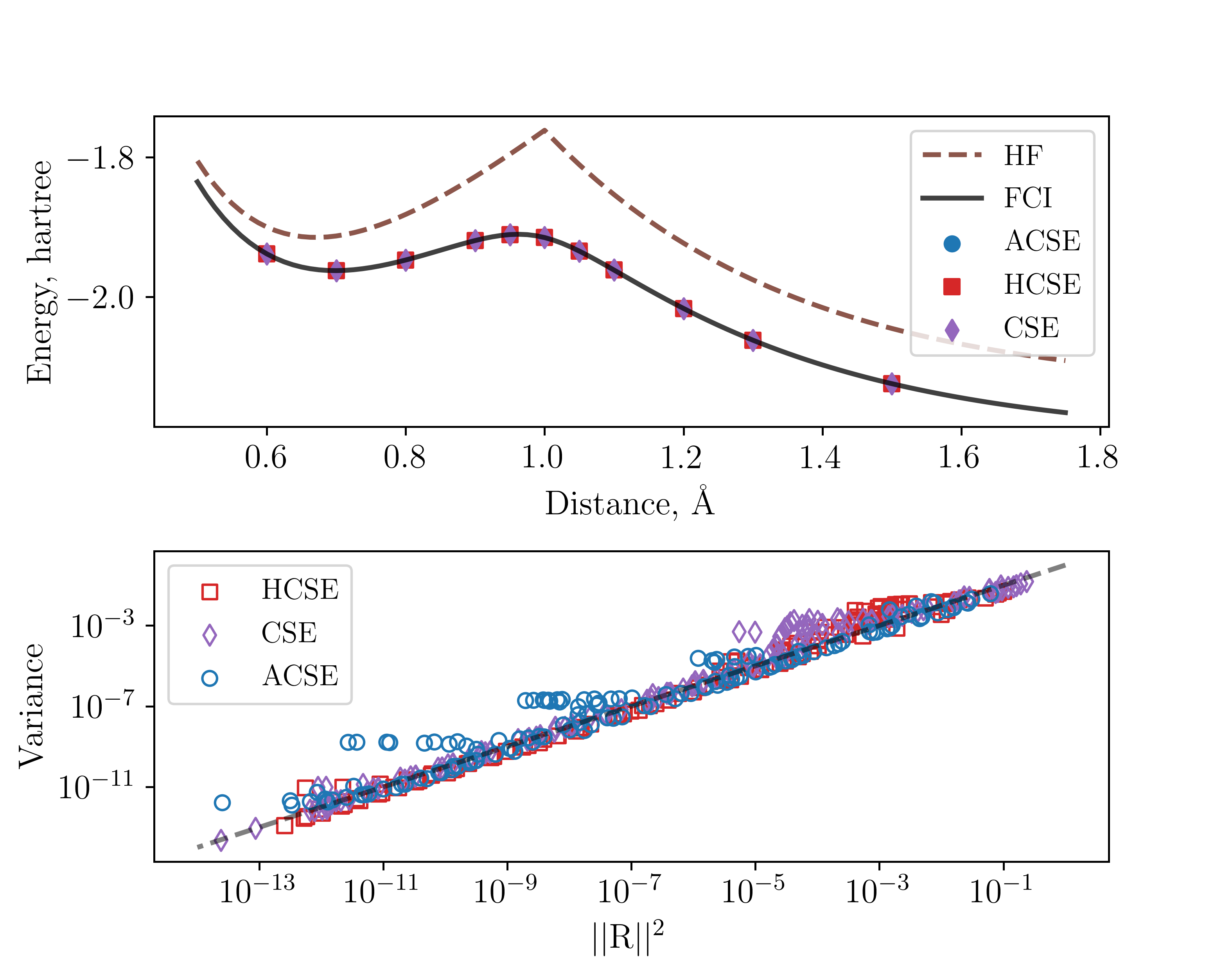}
    \caption{Simulations of rectangular H$_4$ dissociation (with pairs kept at 1\AA) with a noise-free CQE: (a) energy dissociation curves and (b) comparison of CSE, HCSE, and ACSE residual norms and energy variances under several points in trajectories taken from (a).}
    \label{fig:h4_simulations}
\end{figure}

Figure~\ref{fig:h4_simulations} shows the performance of three CQE methods---CSE, ACSE, and HCSE---on noiseless simulations in the stretching of rectangular H$_4$ along one of its sides. The energy surface possesses a discontinuity at the square configuration that is difficult to capture with single-reference methods~\cite{Finley1995,Romero2017}.  With all methods we are able to obtain the FCI solution at different dissociation lengths including the discontinuity with its multi-reference correlation.  In (c) we can also see a close relation between the (squared) norm of the residuals and the energy variance.  While the CSE and HCSE residuals linearly follow the variance, the ACSE residual encounters temporary plateaus where it becomes disproportionately small relative to the variance, reflecting that unlike the case with the CSE and HCSE, the solution of the ACSE does not rigorously imply the solution of the Schr{\"o}dinger equation, which is equivalent to the vanishing of the energy variance. Note that an upper bound on the variance in relation to the norms can be derived via the Cauchy-Schwarz inequality.

\begin{figure}
    \centering
    \includegraphics[scale=0.56] {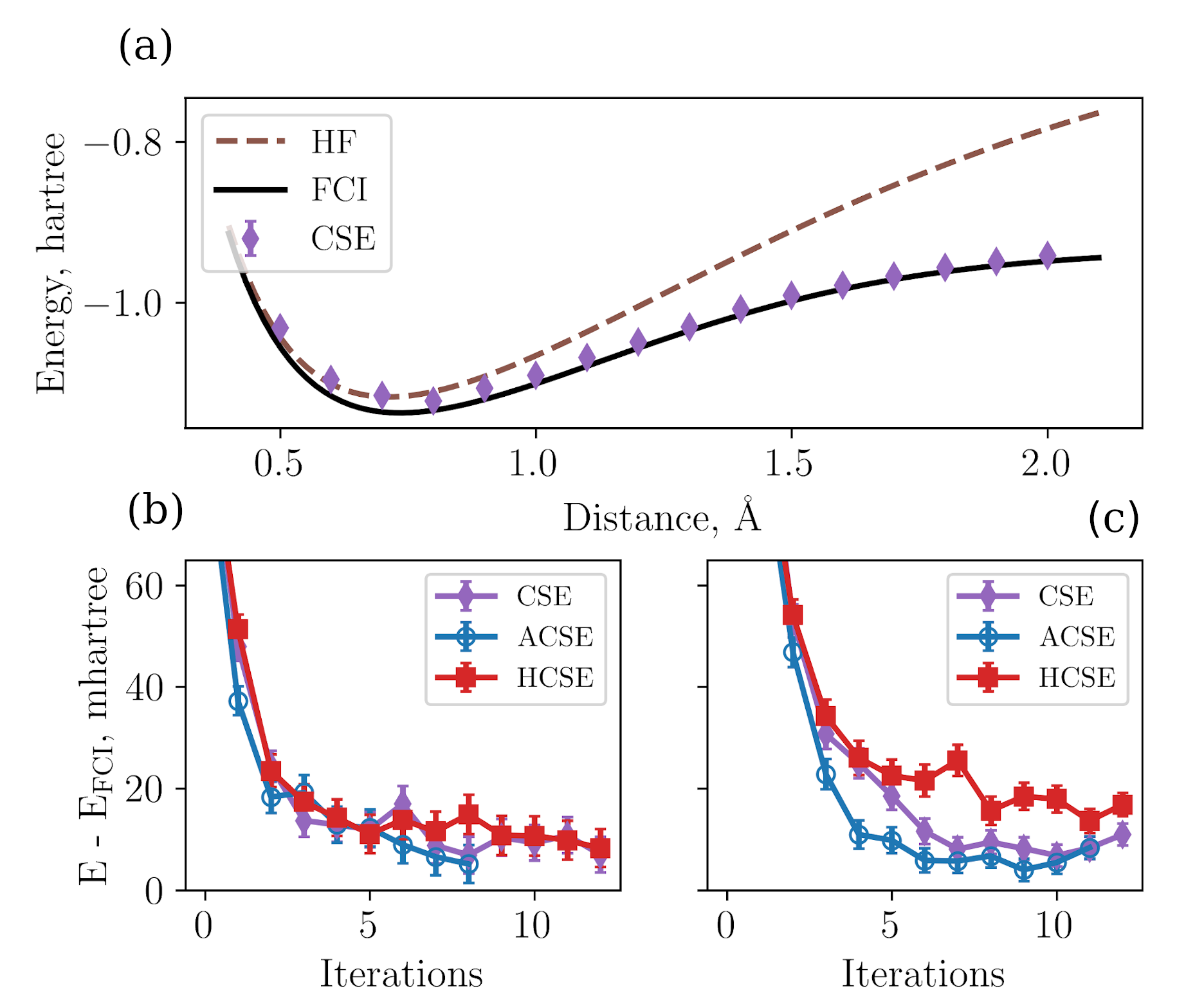}
    \caption{Quantum computations on ``ibm\_lagos" of the dissociation of H$_{2}$ by the CSE, ACSE, and HCSE, as well as classical computations by Hartree-Fock and full Configuration Interaction.}
    \label{fig:h2_lagos}
\end{figure}

Finally, using the IBMQ Quantum Experience, we simulate without error mitigation the dissociation of ${\rm H}_2$ using the CSE.  The simulation requires two qubits---one for the symmetry-tapered wave function and one for the non-unitary dilation. The target wave function is represent by a single qubit using symmetry tapering. Figure \ref{fig:h2_lagos} shows that like the ACSE, the CSE recovers correlation energy across the dissociation surface, with the accuracy limited only by sampling and noise-related errors. The differences in implementation here on noisy devices favors the ACSE, which relative to the CSE requires fewer multi-qubit gates in a more compact ansatz.

\section{Comparisons to Other Eigensolvers}

Many-particle quantum systems, we have shown, can be solved from an exact solution of the CSE on quantum devices. The solution of the CSE is particularly attractive because it has the following important property: a wave function satisfies the CSE if and only if it satisfies the Schr{\"o}dinger equation.  Therefore, the CSE provides a non-exponential-scaling, non-variational ansatz for the wave function without sacrificing exactness relative to the variational solution of the Schr{\"o}dinger equation.

Previously, we have shown that the anti-Hermitian part of the CSE, known as the ACSE, can be solved on quantum devices~\cite{Smart2021_PRL}. While the ACSE ansatz for the wave function has generally been observed to converge to the result from exact diagonalization (i.e. full configuration interaction)~\cite{Smart2022_JCTC,Mazziotti2007}, only the CSE has a mathematical guarantee that its wave function solutions have a one-to-one mapping with the solutions of the Schr{\"o}dinger equation~\cite{Mazziotti2004, Mazziotti2020}.  As discussed in this Article, unlike the ACSE, optimization of the CSE on quantum devices requires the performance of non-unitary transformations. Because the ACSE is a subset of the CSE, the performance of non-unitary transformations that aim to satisfy the CSE can be interwoven with unitary transformations that aim to satisfy the ACSE. Moreover, the balance of unitary and non-unitary transformations can be controlled to optimize both accuracy and convergence on noisy intermediate-scale or fault-tolerant quantum devices, possibly aided through compression of two-body operators~\cite{Schwerdtfeger2012,Hoy2015,Rubin2022}.  The CSE can be invoked at the end of a CQE based on the ACSE or a VQE calculation to refine or verify convergence to an exact stationary point of the Schr{\"o}dinger equation.

The solution of the CSE can be viewed as solving the many-particle problem on quantum devices based on an integration or contraction of the many-particle state, which in the context of the ACSE has been called a contracted quantum eigensolver (CQE).  In contrast to the variational quantum eigensolver (VQE)~\cite{Peruzzo2014, Tilly2021} which generally has an approximate wave function ansatz and relies upon variational improvement, the CQE based on the CSE has a scalable wave function ansatz that by definition converges to the exact solution of the Schr{\"o}dinger equation.  Furthermore, unlike adaptive VQE~\cite{Grimsley2018}, the CQE does not require variational re-optimization of all parameters at each iteration and is guaranteed to be exact when its energy gradient vanishes.  Finally, the CSE-based CQE provides a framework for exploring non-unitary transformations in the quantum simulation of electronic structure. The combination of these unique features and potential advantages show that the exact solution of the CSE on quantum devices provides an important step forward for the quantum simulation of many-particle systems on intermediate-term and future quantum computers.

\begin{acknowledgments}
D.A.M. gratefully acknowledges the Department of Energy, Office of Basic Energy Sciences, Grant DE-SC0019215 and the U.S. National Science Foundation Grants No. CHE-2155082 and No. CHE-2035876.  We acknowledge the use of IBM Quantum services for this work. The views expressed are those of the authors, and do not reflect the official policy or position of IBM or the IBM Quantum team.
\end{acknowledgments}

\section{Methods}

The paired spin model, used in Figure~\ref{fig:fig1_acse_cse_comp}, describes a three-state system of sequential pair excitations, where the potential-free eigenstates are separated by double excitations, and the ground to second excited state requires a quadruple excitation. Denoting a pair population by $1$, we can map this to the following states:
\begin{align}
|0\rangle &= |1010\rangle \\
|1\rangle &= |1001\rangle \\
|2\rangle &= |0110 \rangle \\
|3\rangle &= |0101\rangle
\end{align}
Under a symmetric potential states $|1\rangle $ and $|2\rangle$ are degenerate, with the proper eigenstate being the positive linear combination of the two. Thus, we can map the three states, $|0\rangle$, $\frac{1}{\sqrt{2}}(|1\rangle + |2\rangle)$, and $|3\rangle$, to the unit sphere. Importantly, $|0\rangle \leftrightarrow |1\rangle$ and $|1\rangle \leftrightarrow |2\rangle $ are mediated through a single pair excitation (i.e. a double excitation), whereas the $|0\rangle \leftrightarrow |2\rangle$ transition requires two pair excitations, i.e. a four-electron term. This situation also can occur between combinations of high and low-spin states, which are not coupled through two-electron Hamiltonians \cite{Valdemoro2011}.

Results in Figs.~\ref{fig:h2_lagos} and \ref{fig:h4_simulations} were performed using the \textsc{hqca} (v22.9)\cite{Smart_hqca_-_hybrid} set of tools, which utilizes \textsc{qiskit} (v0.29.0)\cite{Qiskit}, IBM Quantum\cite{Quantum2022}, and \textsc{pyscf} (v1.7.6)\cite{Sun2017} for interfacing with quantum simulators and obtaining electron integrals for circuit based simulations. For Fig.~\ref{fig:h2_lagos}, 16000 shots for circuit measurements were used for the CSE and HCSE, and 8000 for the ACSE, and $\mathbb{Z}_2$ symmetries were applied in order to reduce the Hamiltonian to a single qubit.

\noindent {\it Data availability.} Data will be made available upon reasonable request.

\noindent {\it Code availability.} Code will be made available on a public Github repository upon publication.

\bibliography{
    biblio/books.bib,
    biblio/qc.bib,
    biblio/cqe.bib,
    biblio/cse.bib,
    biblio/nisq.bib,
    biblio/misc.bib,
    biblio/nonunitary.bib,
    biblio/software.bib
    }
\end{document}